\documentclass[reprint,prl,prd]{revtex4-1}

\usepackage{amsmath}
\usepackage{amssymb}
\usepackage{graphicx}
\usepackage{dsfont}
\usepackage{float}
\usepackage{braket}
\usepackage{slashed}
\usepackage{hyperref}
\usepackage{enumerate}

\begin{document}
%opening
\title{Collisional interactions between self-interacting non-relativistic boson stars: effective potential analysis and numerical simulations}
\author{Eric \surname{Cotner}}
\affiliation{Department of Physics and Astronomy, University of California, Los Angeles, CA, 90095-1547, USA}
\email{ecotner@physics.ucla.edu}

\begin{abstract}
Scalar particles are a common prediction of many beyond the Standard Model theories. If they are light and cold enough, there is a possibility they may form Bose-Einstein condensates, which will then become gravitationally bound. These boson stars are solitonic solutions to the Einstein-Klein-Gordon equations, but may be approximated in the non-relativistic regime with a coupled Schr\"odinger-Poisson system. General properties of single soliton states are derived, including the possibility of quartic self-interactions. Binary collisions between two solitons are then studied, and the effects of different mass ratios, relative phases, self-couplings, and separation distances are characterized, leading to an easy conceptual understanding of how these parameters affect the collision outcome in terms of conservation of energy. Applications to dark matter are discussed.
\end{abstract}

\keywords{boson star, Bose star, Q-ball, self-interaction, dark matter, scattering}
\pacs{95.35.+d, 03.65.Ge, 03.75.Nt, 02.70.Bf, 03.65.Sq}

\maketitle

\section{Introduction}
One of the outstanding problems facing modern astrophysics and particle physics is the composition of the dark matter which makes up a large fraction of the energy density of the Universe. In the quest to identify the characteristics which define this type of matter, many authors have considered numerous models. One of the main candidates in this quest is the WIMP, a class of particle with GeV- or TeV-scale mass which (as the name suggests), interacts weakly with itself and/or normal baryonic matter. Though WIMP models in the $\Lambda$CDM paradigm can reproduce the observed large-scale structure of the Universe, there is some tension between $\Lambda$CDM simulations and observation on galaxy-sized scales. Specifically, there are issues concerning how sharply peaked the density profiles of DM-dominated galaxies are \cite{wysegilmore07}, and the abundance and luminosity of satellite galaxies \cite{bullock10}. Simulations predict singular behavior near the center of a typical galaxy, while observations of dwarf spheroidal galaxies prefer a smoother, core-like profile. Propositions to explain this discrepancy have invoked processes such as baryonic feedback \cite{pontzen14}, in which supernovae or other baryonic astrophysical phenomena near the galactic nucleus push dark matter outward, and self-interacting dark matter (SIDM) \cite{elbert14}, in which infalling particles can transfer momentum to those in the core, smoothing the density profile. However, baryonic interaction with dark matter and the SIDM cross-section are highly constrained by measurements from the Bullet Cluster \cite{randall07}, direct-detection experiments \cite{albuq13}, and black hole growth \cite{hennawi01}. \newline
Indeed, a DM candidate exists which can naturally explain the cusp-core problem while simultaneously reproducing the same cosmological-scale structure of the Universe in the same fashion as $\Lambda$CDM \cite{schiveChiueh14}, and may even be able to resolve the mysteries surrounding the collisions of the Bullet Cluster and Abell 520 \cite{leelim08}. A boson star, or soliton composed of a self-gravitating Bose-Einstein condensate, has a series of interesting properties which make it a good candidate to resolve some of the outstanding issues inherent with $\Lambda$CDM on small scales. The component particles making up a boson star can, to good approximation, share a macroscopic wave function due to their Bose-Einstein statistics. Just as a localized particle's wave function naturally spreads with time, a boson star will expand until the attractive force of gravity balances the outward ``quantum pressure", leading to a stable solitonic state. The shape of the resulting density profile is devoid of a singularity, and is smooth and continuous at the origin. In addition, collisions between two boson stars may cause the two to either stick together, scatter inelastically, or pass right through each other (depending on a number of factors), leading to momentum transfer, agglomeration/fracturing of compact objects, or spreading of the central density cusps predicted in $\Lambda$CDM simulations. This effect exists regardless of whether or not the scalar field has significant self-interactions in its potential.\newline
There are many theoretically-motivated candidates to make up the scalar field in question. The axion, for example, is expected to be an extremely light $(m \lesssim 1\, \mu\text{eV})$ scalar, and readily forms a condensate at relatively high temperatures \cite{guth14}. The existence of the Peccei-Quinn axion solves the strong CP problem, and axion-like fields are a general prediction of string theories \cite{svrcek06}. Scalar superpartners of the fermionic fields of the Standard Model, though expected to be relatively heavy, may be good candidates, and have been studied extensively under the name SUSY Q-balls \cite{kusenko05}\cite{kusenko01} (where the Q refers to a conserved Noether charge) in the absence of gravitational interaction. Couplings to the Standard Model Higgs or some other scalar with a Higgs mechanism, if they exist, might also provide a means for creating stable Q-balls from condensed scalar fields. In addition, microlensing experiments from the MACHO \cite{macho00}, EROS \cite{eros03} and OGLE \cite{ogle10} collaborations have detected a significant excess of events over those expected simply from stellar populations, and puts the expected mass of these objects between $0.15 - 0.9 \,M_\odot$ at a 95\% confidence interval. Could the detected compact objects be made up of boson stars?\newline
In this paper, I will outline the existence and classification of stable, non-relativistic boson stars, and will derive approximate analytical profiles and properties of these stars in section \ref{StableStates}. I then focus on the binary interactions between two such stars, and the different head-on scattering outcomes (effectively one-dimensional collisions with zero angular momentum) based on initial velocity, distance, relative phase and mass, and the degree of self-coupling in section \ref{BinaryInteractions}. These findings are then verified through numerical scattering simulations in section \ref{NumericalSims}. Finally I end with a discussion of the possible application of these results to dark matter phenomenology and future research in section \ref{Discussion}.

\section{Existence and stability of boson stars in the non-relativistic limit} \label{StableStates}
We begin with the action for a scalar field coupled with gravity:
\begin{gather}
S = \int d^4x\, \sqrt{-g} \left[ \frac{1}{2\kappa} R + \nabla_\mu \varphi^\dagger \nabla^\mu \varphi - V(\varphi) \right]
\end{gather}
In the non-relativistic approximation, we can write $\varphi$ in terms of a complex wave function by factoring out the harmonic time dependence due to the rest mass:
\begin{gather}
\begin{cases}
\varphi = \frac{1}{\sqrt{2m}} \left( e^{-imt}\psi + e^{imt}\psi^* \right) & \varphi^* = \varphi \\
\varphi = \frac{1}{\sqrt{2m}} e^{-imt}\psi & \varphi^* \not= \varphi
\end{cases}
\end{gather}
We may neglect terms containing an exponential factor since they will average out to zero due to rapid oscillation of the mass frequency, and use the weak-field gravity ansatz $g^{00} = 1 + 2\phi$, $g^{ij} = -(1 +2 \phi)$, $g^{i0} = g^{0j} = 0$. Variation of the action then leads to the Schr\"odinger-Poisson system for a self-interacting scalar field:
\begin{gather}
i\dot{\psi} = -\frac{1}{2m} \nabla^2\psi + \frac{\lambda}{8m^2} |\psi|^2 \psi + m\phi \psi \\
\nabla^2\phi = 4\pi Gm |\psi|^2
\end{gather}
where $\psi$ is the bosonic wave function, $\phi$ is the gravitational potential, and $\lambda$ is the coupling constant due to a quartic self-interaction: $V(\varphi) = \frac{\lambda}{4} |\varphi|^4$. Higher-order effective self-couplings may exist in principle, but we shall neglect them here. Higher-order self-couplings can lead to the formation of solitonic states even in the absence of gravity (a good review of this may be found in Lee and Pang \cite{leepang92}), but a quartic interaction by itself is not enough. These types of field configurations are referred to as mini-boson stars.\newline
In the Hartree-Fock approximation, we may assume that the entire collection of particles share a single wave function, as they have formed a Bose-Einstein condensate. The critical temperature for a non-relativistic condensate is given by \cite{pathria96}:
\begin{align}
k T_c &= \frac{2\pi}{m} \left(\frac{n}{\zeta(3/2)}\right)^{2/3} \\
&= 1.59\text{ MeV} \left(\frac{m}{10^{-9}\text{ eV}}\right)^{-5/3} \left(\frac{\Omega}{.25}\right)^{2/3} \nonumber
\end{align}
where we have substituted values for a typical uniformly-distributed axion-mass dark matter particle under the assumption it makes up 100\% of the DM. The temperature of such a particle today is surely below this limit, and in regions of higher density (such as in galaxies and solitons), this critical temperature will be even higher. \newline
We may find approximate stable ground states through the use of the variational method. First, using the Green's solution for the gravitational potential $\phi$, we may calculate the expectation value of the Hamiltonian in an arbitrary state $\ket{\psi}$ in the following manner \cite{guth14}:
\begin{widetext}
\begin{gather}
\braket{H} = \frac{1}{2m}\int d^3x\, |\nabla \psi|^2 + \frac{\lambda}{16m^2} \int d^3x\, |\psi|^4 - \frac{Gm^2}{2} \int d^3x \int d^3x'\, \frac{|\psi(\vec{x})|^2 |\psi(\vec{x}')|^2}{|\vec{x} - \vec{x}'|}
\end{gather}
\end{widetext}
A good first guess for a solitonic ground state may be a Gaussian profile:
\begin{gather}
\psi(\vec{x}) = \left(\frac{2N^{2/3}k^2}{\pi}\right)^{3/4} e^{-k^2r^2}
\end{gather}
where $k$ is an inverse length-scale variational parameter or wave vector, and $\ket{\psi}$ is properly normalized such that $\braket{\psi|\psi} = N$. Substitution of this state into the expectation value for the Hamiltonian leads to the equation $\braket{H} = \frac{3N}{2m}k^2 + \frac{\lambda N^2}{16\pi^{3/2}m^2}k^3 - \frac{Gm^2N^2}{\pi^{1/2}}k$, and variation with respect to $1/k$ returns
\begin{gather}
1/k = \frac{3\pi^{1/2}}{2} \frac{1}{Gm^3N} \left( 1 + \sqrt{1 + \frac{1}{12\pi^2} \lambda Gm^2N^2} \right)
\end{gather}
This is the same result derived by Chavanis \cite{chavanis11}, and the single-soliton analysis that follows is nearly identical. He also uses a similar ``effective potential" formalism for the radius of the boson star, though he does not extend it to interactions of solitons as is done in this paper. Since dark matter is expected to be weakly self-interacting, it is useful to look at the weak interaction limit where $\xi \equiv \lambda Gm^2N^2/12\pi^2 \ll 1$. (More precisely, $|\xi|\ll 1$, since there is no restriction on the sign of $\lambda$ if we invoke higher-order couplings to prevent the Hamiltonian from being unbounded from below.)
\begin{gather}
1/k \approx 3\pi^{1/2} \frac{1}{Gm^3N} \left( 1 + \frac{1}{48\pi^2} \lambda Gm^2N^2 \right)
\end{gather}
In the case of no interactions whatsoever, we have
\begin{align}
1/k &\approx \frac{3\pi^{1/2}}{Gm^3 N} \approx \frac{3 \pi^{1/2}}{G m^2 M} \\
&= 0.88 \left(\frac{m}{10^{-9} \text{ eV}}\right)^{-2} \left(\frac{M}{1 \text{ M}_\odot}\right)^{-1} \text{km} \nonumber
\end{align}
Since the radius scales as $M^{-1}$, this means that more massive condensates are more tightly gravitationally bound and have a smaller spatial scale. For the parameters given above, this condensate is incredibly tiny given its mass; so tiny that it's actually within its Schwarzschild radius, meaning it would have collapsed to a black hole at this mass (see maximum mass in eqn. \ref{eq:2}). However, even a tiny repulsive self-coupling $\lambda$ can get around this. It is important to note that the $|\xi| \ll 1$ limit does not necessarily imply $|\lambda| \ll 1$; the particle number of the condensate must also be small enough to satisfy this inequality. Likewise, $\xi \gg 1$ does not imply $\lambda \gg 1$. This means the self-interaction parameter may be very large for large particle number, even if the self-coupling $\lambda$ itself is quite modest. When $\xi \gg 1$, the soliton becomes more diffuse and a Gaussian wave function is no longer a good approximation to its shape. However, the length scale parameter should be within a factor of order unity of the real thing, so we can still glean some information from the Gaussian variational wave function in this regime. In this limit, the length scale parameter is
\begin{align}
1/k &\approx \frac{3\pi^{1/2}}{2} \frac{1}{G m^3 N} \left(\xi^{1/2} + 1 + 2\xi^{-1/2}\right) \\
&\approx 2.4 \times 10^{17} \left(\frac{m}{10^{-9} \text{eV}}\right)^{-2} \left(\frac{\lambda}{10^{-6}}\right)^{1/2} \text{kpc} \nonumber
\end{align}
where the final expression is in the limit $N \rightarrow\infty$. For this set of parameters (corresponding to $\xi \sim 10^{23}$), the condensate is \textit{enormous}, stretching beyond the observable universe! Of course this is an extreme example, but I just wanted to show the enormous effect even a tiny coupling can have. We can see that the leading order term in this approximation is independent of the particle number, suggesting that in the limit $N \rightarrow \infty$, the spatial extent of such a soliton will approach a finite size. This may have interesting consequences for black hole formation, since even a tiny self-coupling may prevent black hole collapse. We may approximate the critical mass and rough lower bound at which general relativistic effects take hold and black hole collapse may occur by comparing the Schwarzschild radius to the soliton radius $R_s = 2GM \sim 1/k$ which implies:
\begin{align}
M_\text{max} &\sim \sqrt{\frac{3\pi^{1/2}}{2G^2 m^2}} \approx 10^{-1} \left(\frac{m}{10^{-9} \text{ eV}}\right)^{-1} M_\odot \label{eq:2}
\intertext{in the $|\xi| \ll 1$ regime, and}
M_\text{max} &\sim \frac{1}{8m^2} \sqrt{\frac{3\lambda}{\pi G^3}} \\
&\approx 10^{32} \left(\frac{\lambda}{10^{-6}}\right)^{1/2} \left(\frac{m}{10^{-9} \text{ eV}}\right)^{-2} M_\odot \nonumber
\end{align}
in the $\xi \gg 1$ regime. This agrees with the Kaup limit \cite{kaup68}\cite{ruffini69} in the non-interacting case, and with the analysis of Colpi, Shapiro and Wasserman \cite{colpi86} in the strongly-interacting case. As is evident from this comparison, solitons without self-interaction could potentially readily form many small black holes (especially at higher particle mass), whereas those with self-interaction are stable against gravitational collapse for all practical purposes, unless the self-coupling is extremely small.
The binding energy is given by
\begin{gather}
E_0 = -\frac{8G^2 m^5 N^3 \left( 3 + 2 \xi + 3 \sqrt{1 + \xi} \right)}{36\pi(1 + \sqrt{1 + \xi})^3}
\end{gather}
Which in the two extreme limits are
\begin{gather}
E_0 \approx \begin{cases}
-\frac{G^2 m^5 N^3}{6 \pi }+\frac{\lambda G^3 m^7 N^5}{432 \pi^3} + O(\lambda^2) & |\xi| \ll 1 \\
-\frac{8G^{3/2} m^4 N^2}{3\sqrt{3\lambda}} + \frac{8\pi G m^3 N}{\lambda} + O(\lambda^{-3/2}) & \xi \gg 1
\end{cases}
\end{gather}
which are in good agreement with several numerical analyses \cite{ruffini69}\cite{membrado89}\cite{jetzer92}. The total mass of the star in this state is given by $M = mN + E_0$, which since the binding energy is negative due to the gravitational interaction, is energetically favorable to the state in which the component particles making up the condensate are unbound, ensuring stability against dissolution of the soliton. The soliton is also classically stable against radiation of single particles since $E_0(N-1) - E_0(N) > 0$, meaning that the bound state with $N-1$ particles has a higher energy than the bound state with $N$ particles. Unfortunately, it becomes clear that the non-relativistic approximation breaks down very quickly once the binding energy per particle approaches its rest mass, and therefore is only valid in the regime
\begin{gather}
1 \gg \frac{|E_0|}{mN} \approx \begin{cases}
\frac{G^2 m^4 N^2}{6\pi} & \xi \ll 1 \\
\frac{8 G^{3/2} m^3 N}{3\sqrt{3\lambda}} & \xi \gg 1
\end{cases}
\end{gather}
This analysis therefore only holds for galaxy-mass solitons when the particle mass is very light, $m \lesssim 10^{-21}\text{ eV}$ when there is no self-interaction \cite{schiveChiueh14}, and 30 eV when $\lambda = 10^{-6}$. This upper limit might be relaxed if the dark matter content of the galaxy is instead composed of many smaller solitons. If we assume boson star masses comparable to a solar mass, the particle mass can be much larger before the non-relativistic analysis breaks down, as high as $m \lesssim 10^{-9}\,\text{eV}$ for $\lambda = 0$ and 30 MeV if $\lambda = 10^{-6}$. It bears repeating that these upper limits are by no means actual physical limits, just the limits of applicability of this analysis, and a fully relativistic model must be used for masses beyond this. \newline
As should be obvious, the binding energy becomes complex when $\lambda < -12\pi^2/2Gm^2N^2$. From the form of the Schr\"odinger equation, we can see that the self-interaction term contributes energy positively when the density rises; therefore it represents a short-range self-repulsion when $\lambda$ is positive. When it is negative, this is now a short range attraction. The parameter range where the energy becomes complex thereby signifies an instability where the combined attractive force of gravity and self-interaction causes the soliton to either collapse, or split into multiple smaller solitons until the binding energy of each is no longer complex. This critical mass occurs at
\begin{gather}
M_\text{max} \approx \frac{3\pi}{\sqrt{2G|\lambda|}} = 6.7 \times 10^3 \left(\frac{|\lambda|}{10^{-6}}\right)^{-1/2} \text{M}_P
\end{gather}
which is in general agreement with the work of Eby, Kouvaris, Nielsen, and Wijewardhana \cite{eby15}. It is unclear from this variational analysis which of these situations would occur, but the results of numerical simulations suggest a combination of both.

\section{Head-on binary interactions} \label{BinaryInteractions}
Once the stable states have been found, the next question is how do two or more of these solitons interact with each other. Do they stick, recoil, or pass right through each other during a collision? We specialize to the case of head-on scattering (effectively one-dimensional) for simplicity, though it is simple to add the effects of nonzero angular momentum via addition of an effective potential term ($+J^2/2\mu r^2$ where $\mu$ is the reduced mass) as is commonly done in the solution of radial potential scenarios such as the Kepler problem \cite{jose98}. If we compute the energy of a certain binary configuration, we can use this to answer that question by comparing the energies of different configurations, using this energy as an effective potential for the separation distance. A common method used in undergraduate quantum classes to find the binding energy of the $H_2^+$ molecule will be applicable here. Once again, we use the variational method to find the expectation value of the Hamiltonian, only this time our variational states will be a superposition of two solitons, separated in space, and potentially differing in their relative phase. The separation $d$ will be the variational parameter, and we shall hold the soliton wave numbers $k_i$ constant. In this work, for computational ease, I will suppose that there is no relative motion, so that the two solitons are suspended in their separation. Our superposition state, $\ket{\psi}$, is normalized such that $\braket{\psi|\psi} = N = N_1 + N_2$, so that the total particle number is conserved throughout. This state has the form
\begin{equation} \label{eq:1}
\ket{\Psi(\vec{r})} = A \left[ \ket{\psi(\vec{r} - \vec{d}/2)} + e^{i\alpha} \ket{\psi(\vec{r} + \vec{d}/2)} \right] 
\end{equation}
where $\ket{\psi_i(\vec{r})}$ are the 1-soliton wave functions solved for in the previous section, $\alpha$ is the relative phase between the two solitons, and $A$ is the overall normalization, which is given by
\begin{gather}
A = \sqrt{\frac{N}{N + 2\cos\alpha \braket{\psi_1 | \psi_2}}}, \\
\braket{\psi_1|\psi_2} = 2\sqrt{2 N_1 N_2} \left(\frac{k_1 k_2}{k_1^2 + k_2^2}\right)^{3/2} \exp\left( -\frac{(d k_1 k_2)^2}{k_1^2 + k_2^2} \right)
\end{gather}
Substituting this state into the expectation value of the Hamiltonian, we arrive at a very complicated expression which is better tackled in pieces. There are three terms: the kinetic, self-interaction, and gravitational. The contribution to the kinetic term is of the form
\begin{widetext}
\begin{gather}
E_\text{kin} = \frac{\pi^{3/2} A^2}{8m k_1 k_2} \left[ 3\sqrt{2} (A_2^2 k_1 + A_1^2 k_2) - \frac{16 A_1 A_2 k_1^3 k_2^3}{(k_1^2 + k_2^2)^{7/2}} \left( 2(d k_1 k_2)^2 - 3(k_1^2 + k_2^2) \right) \exp\left(-\frac{(d k_1 k_2)^2}{k_1^2 + k_2^2}\right) \cos \alpha \right]
\end{gather}
\end{widetext}
where $A_1$ and $A_2$ are simply the normalization factors of the single-soliton wave function (e.g. $A_1 = (2N_1^{2/3}k_1^2/\pi)^{3/4}$). We may observe that there is a length scale $\ell = \sqrt{k_1^2 + k_2^2}/k_1 k_2$ which determines a critical separation distance between the two solitons. If we recognize that $R_i \equiv 1/k_i$ is roughly the characteristic radius of the soliton, then we can also understand that $\ell = \sqrt{R_1^2 + R_2^2}$ is the geometric mean of the two radii. Rescaling $d$ so that it is in units of $\ell$ ($x \equiv d/\ell$), we can rewrite the energy as
\begin{widetext}
\begin{align}
E_\text{kin} &= \frac{A^2}{2\sqrt{2} m} \left[ 3\sqrt{2} (N_1 k_1^2 + N_2 k_2^2) - \frac{32\sqrt{N_1 N_2}}{(k_1 k_2)^{3/2} \ell^5} (x^2 - 3/2) e^{-x^2} \cos\alpha \right]
\end{align}
\end{widetext}
After substituting in the variational estimate for the $k_i$ found for the stationary solitons, we find that the kinetic energy is of the form $E_\text{kin} = G^2 m^5 f(N_1, N_2, \alpha, x)$. Thus, the mass of the constituent particles only serves to scale the energy and length factors, and cannot change the features of these curves for fixed particle number. \newline
The phase-dependent bump or well found in the region $x \ll 1$ in the $N_1 \sim N_2$ case is reminiscent of either the nuclear potential found in nuclear scattering and fusion, or the binding energy curve found in atomic physics. When the solitons overlap considerably, this suggests that the two either repel or merge depending on phase. However, the superposition of states we have used to calculate this energy is not likely to hold when $x \ll 1$. This is due to the fact that the nonlinearity introduced by the gravitational interaction violates the superposition principle, so that the superposition of solitons we have started with is technically not allowed. Far from each other, this violation is negligible and assuming superposition is a good approximation. But in the $x \ll 1$ regime, superposition is no longer valid and our trial wave function for individual solitons suffers in accuracy. The solitons should merge, but this description does not account for that. Therefore, treat the results of this analysis in this region with a bit of skepticism. \newline
\begin{figure*}
\centering
\includegraphics[width=1.0\linewidth]{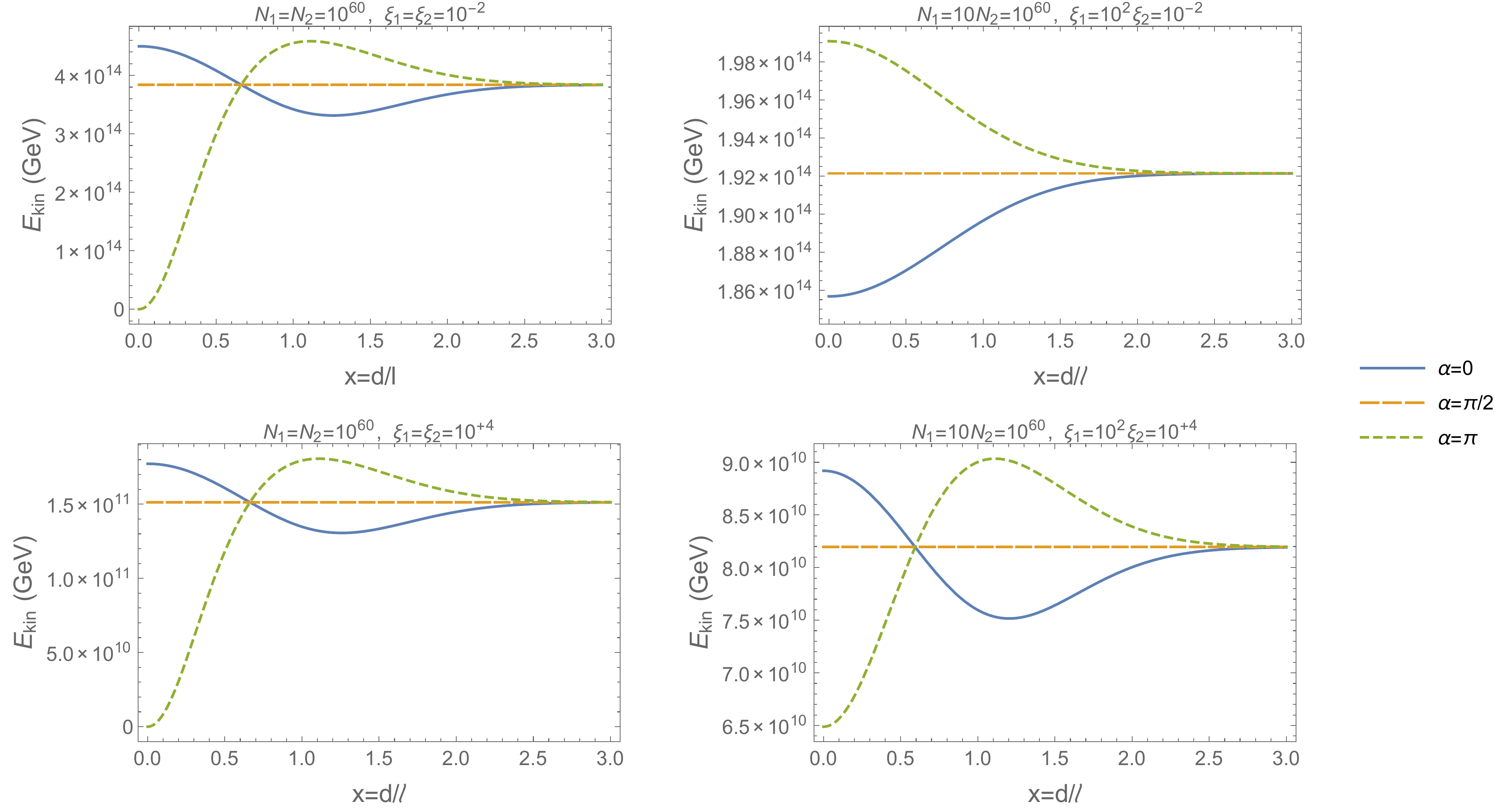}
\caption{Total kinetic energy as a function of scaled separation distance $x=d/\ell$. Solitons of comparable size (left column) have a phase-dependent bump or trough. Binaries where one soliton has much more mass than the other (right column) no longer exhibit this property, unless the self-interaction is very strongly repulsive. Solitons with weak repulsive interactions (top row) have different behaviors in these two regimes, while those with strong repulsive interactions (bottom row) look very similar.}
\label{fig:E_kinetic_plot}
\end{figure*}
When moving along these energy curves slowly enough, the solitons should track the state of lowest energy. If the kinetic contribution to the total energy were the only relevant part, this lowest energy state should be the bottom of the potential well around $x = 1.3$ when $\alpha = 0$, meaning a bound state will form. However, there are still the self- and gravitational interactions to consider, which will change the shape of this curve. \newline
Performing the same procedure as above, we may compute the self-interaction contribution to the energy:
\begin{widetext}
\begin{align}
E_\text{int} &= \frac{\pi^{3/2} \lambda A^4}{64 m^2} \left[ N_1 k_1^3 + N_2 k_2^3 + \frac{32 \sqrt{N_1 N_2}}{(k_1 k_2)^{3/2}} \left( N_1 \left(\frac{k_1}{\ell_1}\right)^3 e^{-3(d/\ell_1)^2} + N_2 \left(\frac{k_2}{\ell_2}\right)^3 e^{-3(d/\ell_2)^2} \right) \cos\alpha \right.\notag\\
&\left.\qquad + \frac{4\sqrt{2} N_1 N_2}{\ell^3} e^{-2(d/\ell)^2} (2+\cos(2\alpha)) \right]
\end{align}
\end{widetext}
There are two more length scales in addition to $\ell$, $\ell_i = \ell \sqrt{1 + 2k_i^2/(k_1^2 + k_2^2)}$. It is clear that if $k_i/k_j \ll 1$, then $\ell_i \rightarrow \ell$, and if $k_i/k_j \gg 1$, then $\ell_i \rightarrow \sqrt{3} \ell$, with intermediate cases falling somewhere in between. \newline
This self-interaction energy is plotted as a function of separation distance in fig. \ref{fig:E_int_plot}.
\begin{figure*}
\centering
\includegraphics[width=1.0\linewidth]{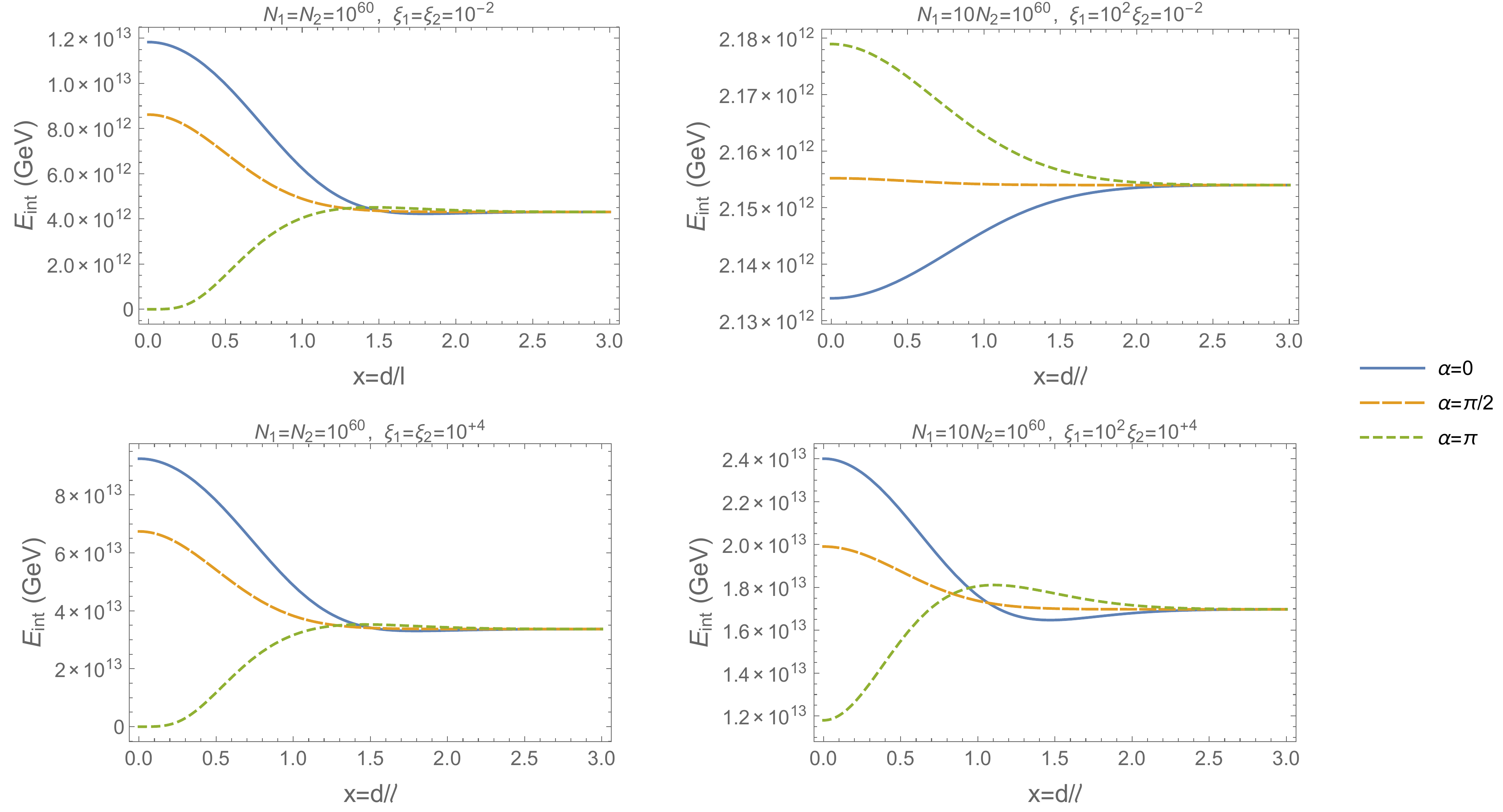}
\caption{Self-interaction energy as a function of scaled separation distance $x = d/\ell$. Solitons of comparable size (left column) exhibit either a repulsive or attractive mutual force depending on phase difference, the shape of which is independent of self-interaction strength. Solitons of asymmetric masses (right column) have a more curious behavior.}
\label{fig:E_int_plot}
\end{figure*}
We may see that in the case of $\lambda > 0$, the mutual force between the two solitons is not necessarily repulsive, and depends on both the relative phase and size of the objects. Specifically, for stars of comparable mass, the interaction is repulsive for wave functions in phase, and attractive for wave functions out of phase. For stars of asymmetric masses, the interaction is attractive for in phase wave functions, and repulsive for out of phase wave functions in the weak self-interaction regime. This behavior is switched in the strong self-interaction regime, where an additional bump/trough forms just as in the kinetic term. In the $\lambda <0$ case, there is no such thing as the strongly-interacting regime, as the individual solitons would be unstable, and behavior is similar to the weak-interaction regime for $\lambda >0$, but with the sign of the energy switched. \newline
The third and final term in the Hamiltonian is the gravitational interaction term. It is composed of four sub-terms which we may break up in the following manner (recalling $\psi = A(\psi_1 + e^{i\alpha} \psi_2)$ and that the integral is invariant under relabeling $\vec{x} \leftrightarrow \vec{x}'$ since the integration regions are the same):
\begin{widetext}
\begin{align}
E_\text{grav} &= -\frac{Gm^2}{2} \int d^3x\, d^3x'\, \frac{|\psi(\vec{x})|^2 |\psi(\vec{x}')|^2}{|\vec{x} - \vec{x}'|} \notag\\
&= -\frac{1}{2} Gm^2 A^4 \int d^3x\, d^3x'\, \left[ \underset{\text{soliton self-energy}}{\underbrace{\psi_1^2(\vec{x}) \psi_1^2(\vec{x}') + \psi_2^2(\vec{x}) \psi_2^2(\vec{x}')}} + 2 \underset{\text{classical gravity}}{\underbrace{\psi_1^2(\vec{x}) \psi_2^2(\vec{x}')}} \right.\notag\\
&\left.\qquad\qquad + 4\cos\alpha\, \underset{\text{long-distance interference}}{\underbrace{\psi_1(\vec{x}) \psi_2(\vec{x}) \left(\psi_1^2(\vec{x}') + \psi_2^2(\vec{x}')\right)}}  + 4\cos^2\alpha\, \underset{\text{short-distance interference}}{\underbrace{\psi_1(\vec{x}) \psi_2(\vec{x}) \psi_1(\vec{x}') \psi_2(\vec{x}')}} \right]/|\vec{x} - \vec{x}'|
\end{align}
\end{widetext}
The self-energy term is simply the binding energies of the individual solitons, and can be coordinate transformed so that it is spherically symmetric and is easily solved:
\begin{gather}
I_1 = \sqrt{2} \pi^{5/2} \left( \frac{A_1^4}{k_1^5} + \frac{A_2^4}{k_2^5} \right) = 8 \sqrt{\frac{2}{\pi}} \left( N_1^2 k_1 + N_2^2 k_2 \right)
\end{gather}
This term, along with the ``classical gravity" term are what you would expect if we were using regular mass densities $\rho = \rho_1 + \rho_2 = \psi_1^2 + \psi_2^2$. However, since we are using wave functions, we must add them together, then square, introducing interference effects. The ``long distance interference" term is identified as such because it contributes more weight to the integral when $\vec{x}'$ is close to the center of either one of the solitons, whereas the ``short distance interference" term must have $\vec{x}'$ near the center of both solitons to contribute significantly. Besides the first term, this expression is not analytically soluble, but we can make some headway numerically by using the Green function in cylindrical coordinates \cite{arfken}:
\begin{widetext}
\begin{gather}
g(\vec{x},\vec{x}') = \frac{1}{4\pi|\vec{x} - \vec{x}'|} = \frac{1}{2\pi^2} \sum_{m=-\infty}^{\infty} \int_{0}^{\infty} dk\, I_m(k s_<) K_m(k s_>) e^{im(\phi - \phi')} \cos(k(z - z'))
\end{gather}
Due to cylindrical symmetry, only the $m=0$ term contributes to the sum after integration over $\phi, \phi'$, and the integrations over $z,z'$ can be performed analytically, leaving us with a triple integral over $s,s'$ and $k$. This must be evaluated numerically. Transforming to dimensionless units $(\sigma,\sigma',\kappa,x) \equiv (s/\ell,s'/\ell,\ell k,d/\ell)$, the expression to integrate shall be
\begin{align}
E_\text{grav} &= -4\pi^2 G m^2 A^4 \ell^3 \int_{0}^{\infty} d\kappa \int_{0}^{\infty} d\sigma \sigma \left[ \int_{0}^{\sigma} d\sigma'\, \sigma' I_0(\kappa \sigma') K_0(\kappa \sigma) + \int_{\sigma}^{\infty} d\sigma'\, \sigma' I_0(\kappa \sigma) K_0(\kappa \sigma') \right] \notag\\
&\quad\times \left[ \frac{A_1^4}{2k_1^2} e^{-\kappa^2/4 \ell^2 k_1^2} e^{-2\ell^2 k_1^2 (\sigma^2 + \sigma'^2)} + \frac{A_2^4}{2 k_2^2} e^{-\kappa^2/4 \ell^2 k_2^2} e^{-2\ell^2 k_2^2 (\sigma^2 + \sigma'^2)} + \frac{A_1^2 A_2^2}{k_1 k_2} \cos(\kappa x) e^{-\kappa^2/8} e^{-2\ell^2 (k_1^2 \sigma^2 + k_2^2 \sigma'^2)} \right. \notag\\
&\left.\quad + 4 A_1 A_2 \cos\alpha e^{-x^2-\ell^2 (k_1^2 + k_2^2)\sigma^2} \left( \frac{A_1^2 \cos\left(\frac{x \kappa}{\ell^2 k_1^2}\right)}{\sqrt{2} k_1 \sqrt{k_1^2 + k_2^2}} e^{-\frac{\kappa^2 \ell_1^2}{8\ell^4 k_1^2}} e^{-2\ell^2 k_1^2 \sigma'^2} + \frac{A_2^2 \cos\left(\frac{x \kappa}{\ell^2 k_2^2}\right)}{\sqrt{2} k_2 \sqrt{k_1^2 + k_2^2}} e^{-\frac{\kappa^2 \ell_2^2}{8\ell^4 k_2^2}} \right) \right. \notag\\
&\left.\quad + \frac{4A_1^2 A_2^2 \cos^2\alpha}{k_1^2 + k_2^2} e^{-2x^2 - \frac{\kappa^2}{2 \ell^4 k_1^2 k_2^2}} e^{-2\ell^4 k_1^2 k_2^2 (\sigma^2 + \sigma'^2)} \right]
\end{align}
\end{widetext}
\begin{figure*}
\centering
\includegraphics[width=1.0\linewidth]{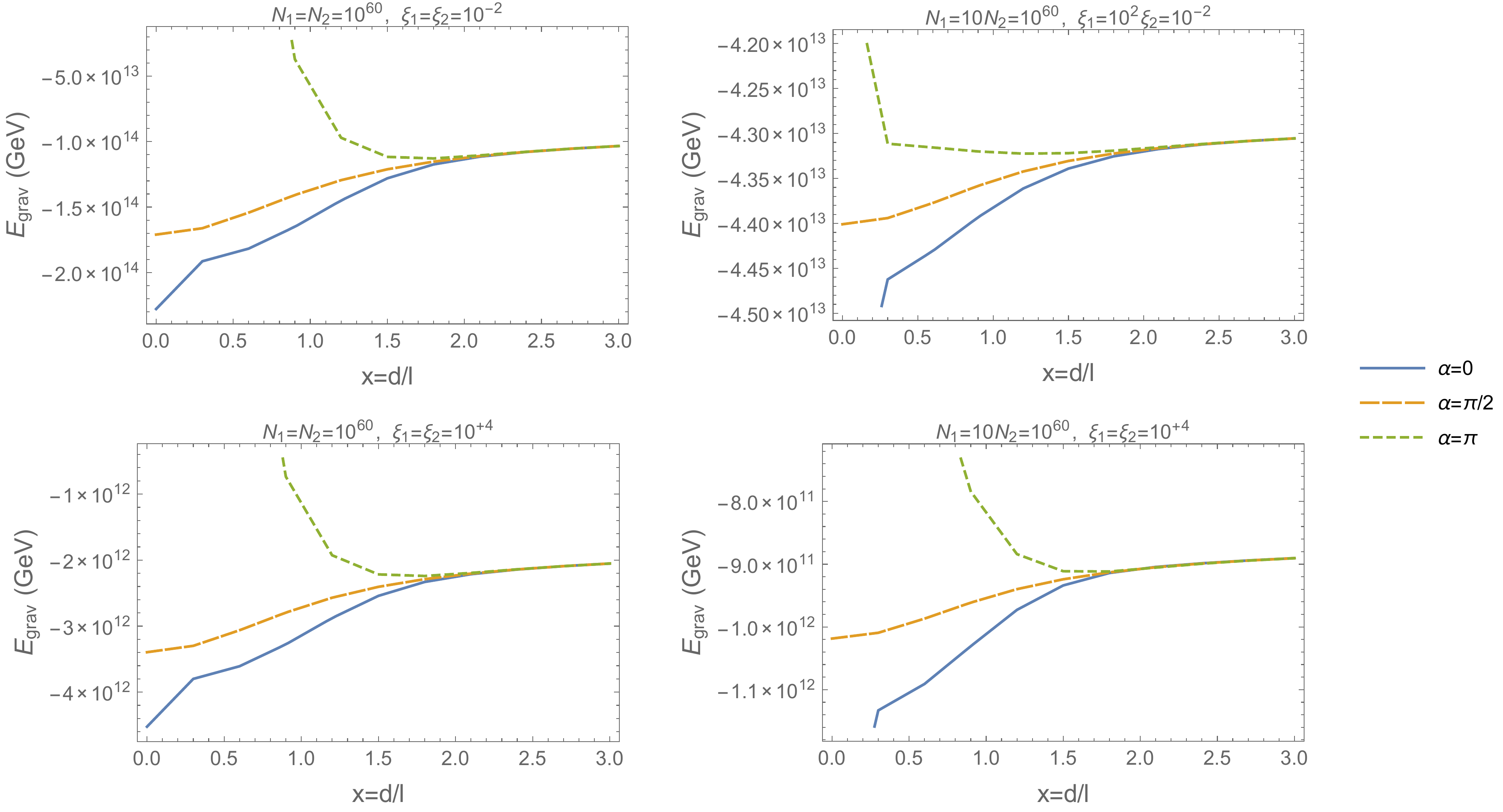}
\caption{Gravitational energy as a function of scaled separation distance $x=d/\ell$. Solitons exhibit a universal behavior of mutual attraction when the relative phase is small, and mutual repulsion when the relative phase is close to $\alpha = \pi$. This repulsive behavior is very strong at small separations except in the case of asymmetric size solitons with weak self-interaction (top right).}
\label{fig:E_grav_plot}
\end{figure*}
Evaluation of this integral for various values of $x$ and $\alpha$ lead to the plots of fig. \ref{fig:E_grav_plot}. We can see that at long distances, the gravitational interaction energy asymptotically approaches $-G M_1 M_2/x - E_1 - E_2$, as one would expect from Newtonian theory. As the distance closes, a prominent rise in energy appears for relative phase $\alpha = \pi$. In the $N_1 \sim N_2$ regime, this rise can even cause the gravitational contribution to become positive (mildly disturbing, but we don't expect this analysis to be valid in the $x \ll 1$ limit as mentioned before), and clearly signals a strongly repulsive interaction. For relative phases less than about $\pi/2$, the energy falls, signaling a mildly attractive force, and potential merger. We can see that in the $\xi \gg 1$ regime, the gravitational binding energy is many orders of magnitude smaller than the equivalent situation with $\xi \ll 1$ due to the internal repulsive force spreading out the soliton. \newline
Having evaluated and discussed the behavior of each of the contributions to the total energy separately, it is now of interest to see what they look like summed together into an effective potential, so that we may grasp the overall behavior of the interaction, which is plotted in fig. \ref{fig:E_tot_plot}.
\begin{figure*}
\centering
\includegraphics[width=1.0\linewidth]{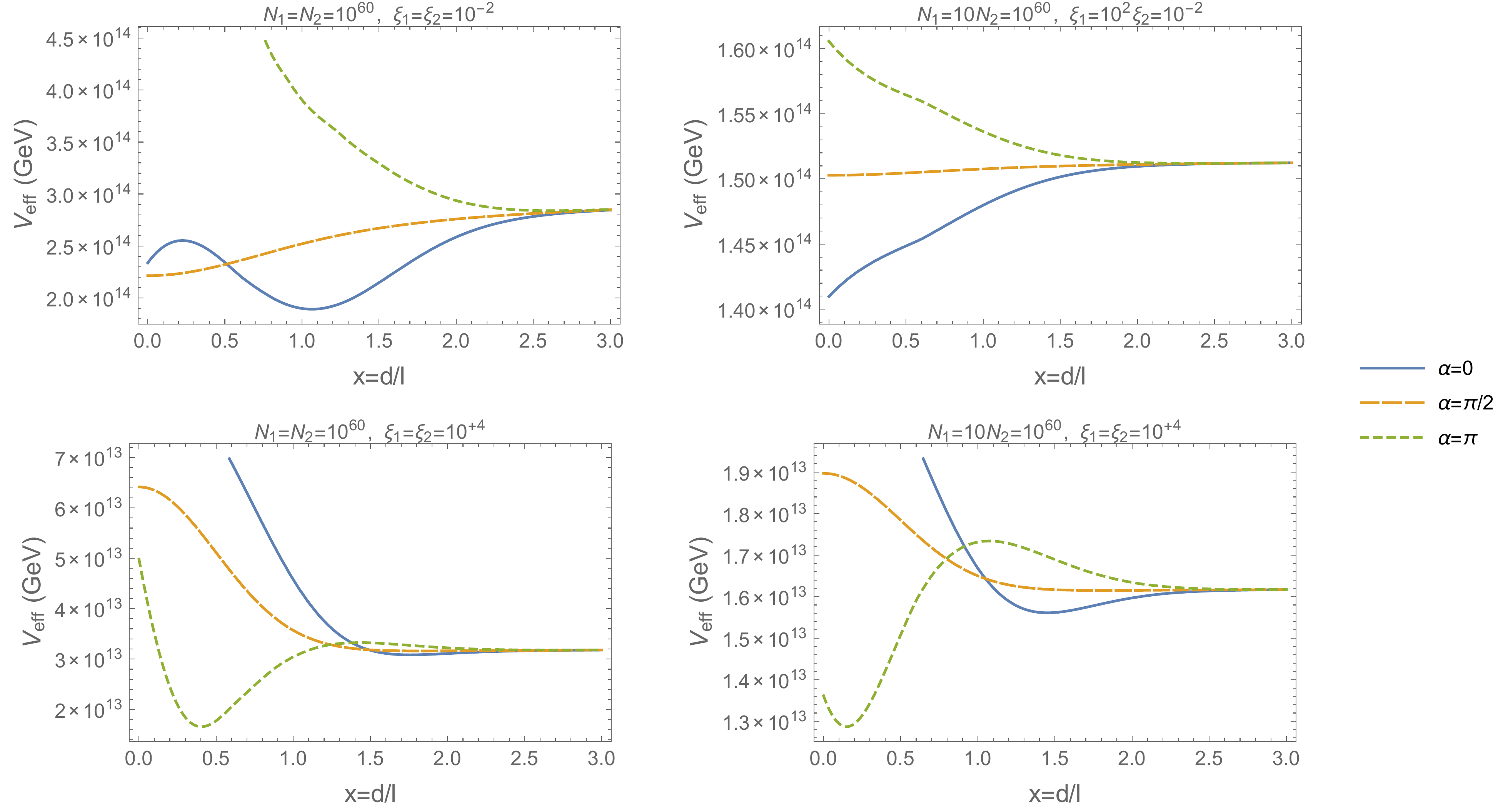}
\caption{Effective potential (total energy) as a function of scaled separation distance $x = d/\ell$. Solitons exhibit a complicated interaction based on the physical parameters, though it is apparent that for $\xi \ll 1$ (top row), solitons in phase have a roughly attractive interaction, while pairs that are out of phase are roughly repulsive. For $\xi \gg 1$ (bottom row), this behavior is switched.}
\label{fig:E_tot_plot}
\end{figure*}
The effective potential curves in each parameter regime are markedly different from each other, though with some unifying overall behavior. First, we may observe that in the $\xi < 1$ regime, in phase solitons are attractive, whereas out of phase solitons are repulsive. In the $\xi > 1$ regime, this is switched. This is because the self-interaction contribution to the energy dwarfs the kinetic and gravitational contributions in this regime, and so the total energy is decently approximated by that term alone. Interesting to note is that in both the ($N_1 \sim N_2$, $\xi \ll 1$) and ($N_1 \gg N_2$, $\xi \gg 1$) regimes, a local minimum appears in the $\alpha = 0$ energy curve, suggesting a permanent bound state may be formed. In the first case, this does not appear likely, as the kinetic energy gained falling into this minimum from infinitely far away should be enough to avoid being bound and make it through the dip to $x=0$, where a merger will occur, unless there is some initial kinetic energy allowing them to escape. In the second case, provided the initial kinetic energy isn't too high, a bound state does appear likely to form since the $\alpha = 0$ curve turns sharply upwards, creating a barrier to merger. The $\alpha = \pi$ curve, however, provides much less resistance to merger due to the much smaller energy barrier. Curiously, since it is more energetically favorable for the two solitons to be out of phase in the $\xi \gg 1$ regime at small $x$, it may be possible to have a shell-like structure where the core of the star is out of phase with respect to the exterior. This would result in a node in the radial wave function, leading one to believe that excited states may easily be formed from collisions of this type.

\section{Numerical Simulation} \label{NumericalSims}
In order to verify the above scattering predictions, I have employed a number of numerical simulations. These simulations solve the time-dependent Schr\"odinger-Poisson system with nonlinear self-interaction for initial states of two solitons, which are then evolved in time to determine scattering behavior. The code utilizes a grid method to calculate the wave function $\psi$ by discretizing Schr\"odinger's equation on a 3D Cartesian grid and using a 2nd-order center-time, 1st-order center-space (CTCS) algorithm ($\partial \psi/\partial t \rightarrow (\psi(t+dt) - \psi(t-dt))/2dt$, $\partial^2\psi/dx_i^2 \rightarrow (\psi(x_i+dx_i) - 2\psi(x_i) + \psi(x_i-dx_i))/dx_i^2$) to iterate the value of the field at each grid point in time. The gravitational potential $\phi$ is also discretized on the same grid, and is solved for using a successive over-relaxation (SOR) algorithm with the $\psi$ field from the previous time step as the source. In order to satisfy the boundary condition that both $\psi$ and $\phi$ should vanish at infinity, I have performed an asymptotic coordinate transformation $\chi_i = \tanh(x_i/R)$ to map the domain from $\mathds{R}^3 \rightarrow [-1,1]^3$, which allows Dirichlet conditions to be imposed on the boundary of the box $[-1,1]^3$. The relation between the probability densities in these two coordinate systems is then
\begin{widetext}
\begin{align}
|\psi(\vec{x})|^2\,d^3x &= |\psi(\vec{\chi})|^2 \left|\frac{\partial(x,y,z)}{\partial(\chi_1,\chi_2,\chi_3)}\right| d^3\chi = \frac{|\psi(\vec{\chi})|^2 R^3}{(1-\chi_1^2)(1-\chi_2^2)(1-\chi_3^2)} d\chi_1\,d\chi_2\,d\chi_3
\end{align}
The parameter $R$ is a scaling factor which is roughly the spatial extent of the radius within which the simulation should be contained. Outside this radius, the simulation will still run correctly, but spatial distances are severely distorted and might not capture all the details of the evolution. In terms of the discretized $\chi$ variables, Schr\"odinger's equation can be solved for the value of $\psi$ at successive time steps:
\begin{align}
\psi_{i_1,i_2,i_3}^{n+1} &= \psi_{i_1,i_2,i_3}^{n-1} + \frac{i\, dt}{(d\chi\, R)^2 m} \left[ (1-\chi_1^2)(1-\chi_1^2 - \chi_1 \, d\chi) \psi_{i_1+1,i_2,i_3}^{n} + (1 - \chi_1^2 + \chi_1\, d\chi) \psi_{i_1-1,i_2,i_3}^{n} - 2 (1 - \chi_1^2) \psi_{i_1,i_2,i_3}^{n} \right] \nonumber \\
&\qquad + (1 \leftrightarrow 2) + (1 \leftrightarrow 3) - \frac{i \lambda\, dt}{4m^2} |\psi_{i_1,i_2,i_3}^{n}|^2 \psi_{i_1,i_2,i_3}^{n} - 2i m\, dt\, \phi_{i_1,i_2,i_3}^{n} \psi_{i_1,i_2,i_3}^{n}
\end{align}
where $n$ denotes the time step and the $i$'s index spatial grid points. $\phi$ refers to the gravitational potential, which is solved using the aforementioned SOR algorithm:
\begin{align}
\phi_{i_1,i_2,i_3}^{n,n_r+1} &= (1-\omega) \phi_{i_1,i_2,i_3}^{n,n_r} + \frac{\omega}{2 \left((1-\chi_1^2)^2 + (1-\chi_2^2)^2 + (1-\chi_3^2)^2\right)} \left[ (1-\chi_1^2) (1-\chi_1^2 - \chi_1 \, d\chi) \phi_{i_1+1,i_2,i_3}^{n,n_r} \right. \nonumber \\
&\left.\qquad + (1-\chi_1^2 + \chi_1\, d\chi) \phi_{i_1-1,i_2,i_3}^{n,n_r+1} + (1 \leftrightarrow 2) + (1 \leftrightarrow 3) - 4\pi G m R^2 \, d\chi^2 |\psi_{i_1,i_2,i_3}^{n}|^2 \right]
\end{align}
where $\omega = 1.9$ is the over-relaxation parameter, and $n_r$ denotes the relaxation time step. This is run repeatedly until the error between steps $(\phi^{n_r+1} - \phi^{n_r})/\phi^{n_r+1} < 10^{-4}$.
\end{widetext}
As initial states, I take a superposition of two stationary states, similar to that of eqn. \ref{eq:1}, though with additional non-constant phase factors to account for initial velocities:
\begin{gather}
\Psi(\vec{r}, t_0) = A \left[ \psi(\vec{r} - \vec{d}/2) e^{i m \vec{v}_1 \cdot \vec{r}} + e^{i\alpha} \psi(\vec{r} + \vec{d}/2) e^{i m \vec{v}_2 \cdot \vec{r}} \right]
\end{gather}
where $\psi$ is the stationary wave function, $m$ is the mass of the scalar particle, $\alpha$ a constant phase factor, and $\vec{v}_i$ are the initial velocities. I do not use initial momentum here because the quantity in the exponential would be the average momentum \textit{per particle}, which is somewhat confusing, especially in scenarios where the two solitons are different masses. In most of the cases that will follow, nonzero initial velocities will be equal and opposite: $\vec{v}_1 = -\vec{v}_2 \equiv \vec{v}$. In order to verify the predictions of the previous sections, I will perform a number of simulations under different conditions, listed here for clarity:
\begin{enumerate}[A.]
\item $N_1 = N_2$, $\xi \ll 1$, $\alpha = 0$, $v_0 = 0$: predicted to fall together, then merge.
\item $N_1 = N_2$, $\xi \ll 1$, $\alpha = 0$, $v_0 \not= 0$: predicted to fall together, then pass through each other mostly intact.
\item $N_1 = N_2$, $\xi \ll 1$, $\alpha = \pi$, $v_0 = 0$: predicted to fall towards each other, reflect, then sit adjacent until eventual merger.
\item $N_1 = N_2$, $\xi \gg 1$, $\alpha = \pi$, $v_0 = 0$: predicted to fall towards each other and merge.
\item $N_1 \gg N_2$, $\xi \ll 1$, $\alpha = \pi$, $v_0 = 0$: predicted to fall towards each other, reflect, then sit adjacent until eventual merger.
\item $N_1 \gg N_2$, $\xi \gg 1$, $\alpha = 0$, $v_0 = 0$: predicted to fall towards each other, reflect, then sit adjacent to each other for an extended period (possibly indefinitely).
\item Single large soliton with self-interaction parameter past critical range $\xi < -1$: predicted to fracture into multiple smaller solitons.
\end{enumerate}
There are too many parameter combinations to cover the entire parameter space, so we will focus on those with possibly interesting effects. In what follows, I will present the results of the simulations in the form of snapshots of the system at relevant times (where $t$ refers to the number of timesteps since the simulation was initiated). The plots are of surfaces of constant $|\psi|^2$, with different contours representing half-logarithmic steps ($c_n = 10^{n/2}$).\newline
The numerical stability of these simulations are governed by three different stability parameters $s,\, s_\lambda,\, s_G$, that determine the numerical instability corresponding to the dynamical, $\lambda \varphi^4$, and gravitational interactions, each of which scales linearly with the time step $\Delta t$. For stability to reign, we need all three parameters to be significantly less than unity. This can always be accomplished by making $\Delta t$ smaller, at the cost of increased CPU time, which scales as $\Delta t^{-1}$ for simulations of the same length of time. This means that for reasonable-length simulations I can run on commercially-available hardware, I constrain $s \sim 10^{-1}$. Unfortunately, the ratio $s_\lambda/s \sim \xi^{1/2}/M^2$ for large $\xi$, where $M$ is the number of grid points along one dimension of the simulation, so that simulating very strongly interacting systems becomes infeasible from a computational standpoint. This forces us to consider simulations of only mildly self-interacting systems with $\xi = 10^1$ rather than the $\xi = 10^4$ systems considered in section \ref{BinaryInteractions}. \newline
In order to verify the convergence properties of this simulation, the code has been run on successively finer meshes of $M^3 = 20^3,\, 40^3,\text{ and } 50^3$ grid points, with comparison between simulation and the effective potential prediction done with the $M=50$ results. Conservation of particle number and total energy have been verified to improve as the mesh is refined. (As a specific example, solitons with radii on the order of the grid spacing would get ``stuck" to a specific grid point, clearly violating energy. This issue is cleared up as $M$ increases.) \newline
Animations of these simulations can be found online at (\url{http://ecotner.bol.ucla.edu/Research/Solitons/BoseStars.html}).

\subsection{$N_1 = N_2$, $\xi \ll 1$, $\alpha = 0$, $v_0 = 0$} \label{ssec:sim1}
In this first simulation, we look at the behavior of two equal-mass boson stars with a negligible self-interaction and no relative phase or initial velocity.
\begin{figure*}
\centering
\includegraphics[width=0.9\linewidth]{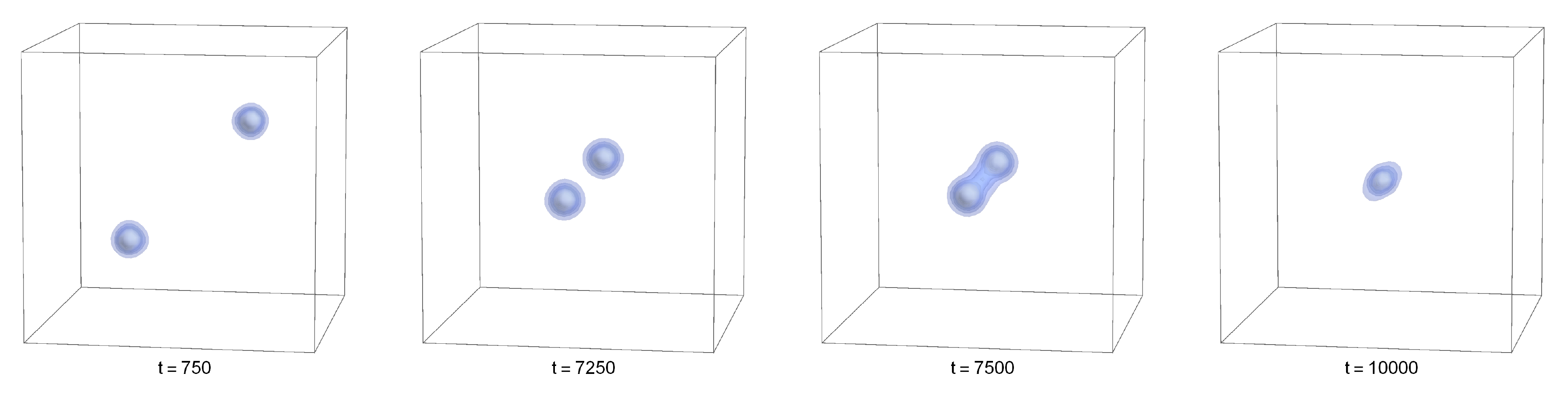}
\caption{Snapshots of two colliding boson stars with $N_1 = N_2 = 10^{60}$, $\xi_1 = \xi_2 = 10^{-2}$, $\alpha = 0$ and $v_0=0$, as explained in section (\ref{ssec:sim1}). Contours are surfaces of constant $|\psi|^2$.}
\label{fig:Sim1}
\end{figure*}
As we can see from fig. \ref{fig:Sim1}, the solitons fall under gravitational attraction toward each other (taking time to accelerate, which explains the large delay between snapshots 1 and 2), merge together, and then sit stationary in the center of mass while experiencing mild radial oscillations which are symmetric under rotations about the axis of approach. This is in line with the prediction from the effective potential: that the system doesn't have enough energy to overcome the hump at the origin and the two solitons merge.

\subsection{$N_1 = N_2$, $\xi \ll 1$, $\alpha = 0$, $v_0 \not= 0$} \label{ssec:sim2}
In this second simulation, we look at the behavior of two equal-mass boson stars with negligible self-interaction and no relative phase. However, they do have initial velocities (directed towards each other).
\begin{figure*}
\centering
\includegraphics[width=0.9\linewidth]{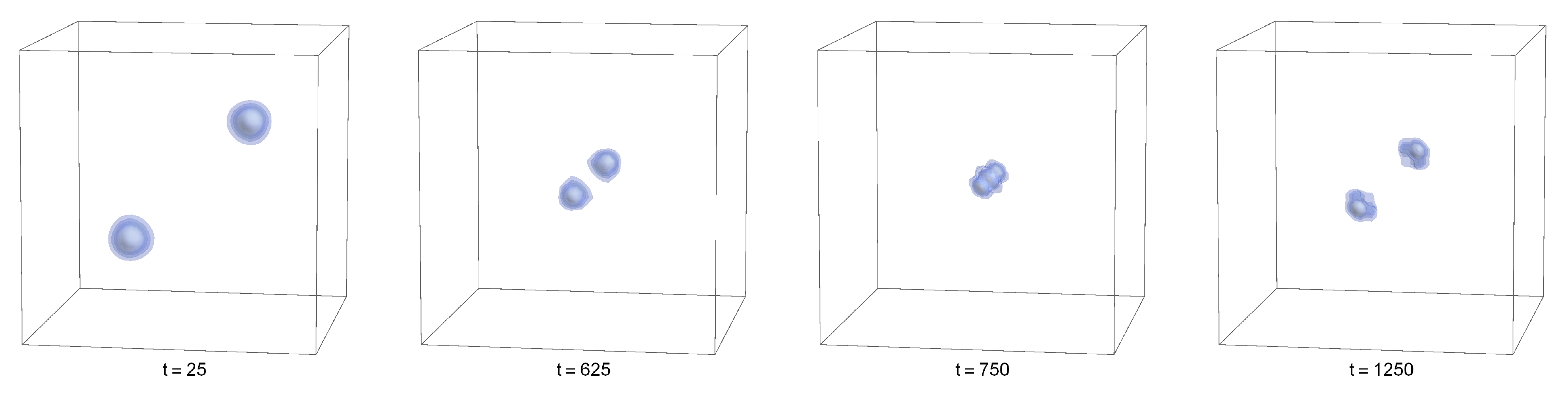}
\caption{Snapshots of two colliding boson stars with $N_1 = N_2 = 10^{60}$, $\xi_1 = \xi_2 = 10^{-2}$, $\alpha = 0$ and $v_0=7 \times 10^{-15}$, explained in section (\ref{ssec:sim2}). Contours are surfaces of constant $|\psi|^2$.}
\label{fig:Sim2}
\end{figure*}
As we can see from fig. \ref{fig:Sim2}, the solitons accelerate towards each other, merge briefly, then continue onwards, their trajectories unaltered. This is as expected, as now the system has enough kinetic energy to get over the hump at the origin of the effective potential. Not shown is that significantly after passing through each other, the solitons bloom outward as though torn apart by the interaction. I believe this to be a numerical artefact, and will check with further simulations. If not, collisions of this type might be a possible mechanism for fracturing of larger boson stars into smaller ones.

\subsection{$N_1 = N_2$, $\xi \ll 1$, $\alpha = \pi$, $v_0 = 0$} \label{ssec:sim3}
This next simulation looks at the effects of initial relative phase differences on the collision. From the effective potential, it appears that a collision between two out-of-phase solitons will result in a repulsive force at close range.
\begin{figure*}
\centering
\includegraphics[width=0.9\linewidth]{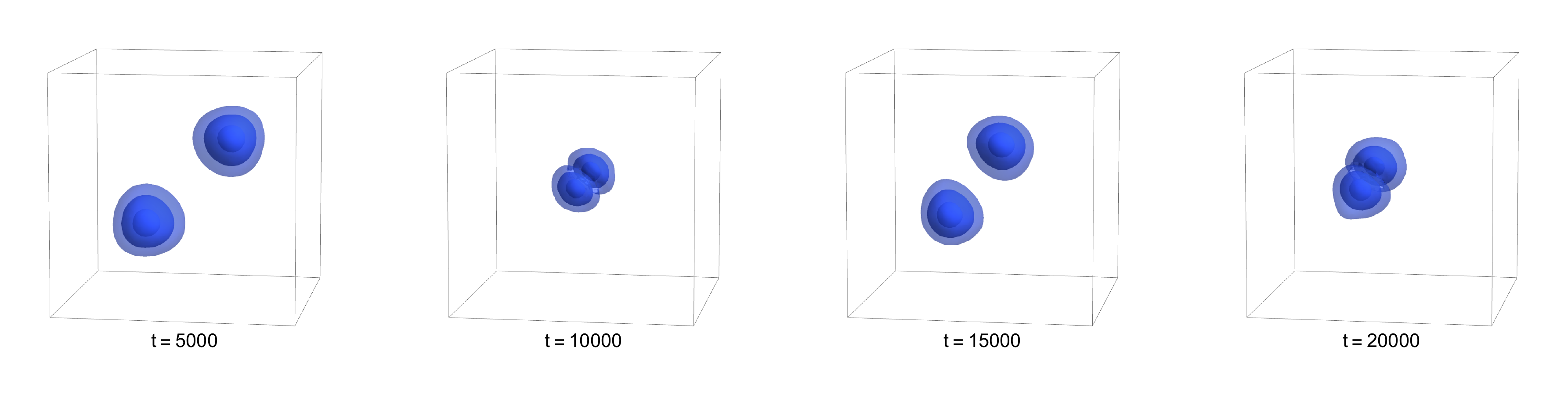}
\caption{Snapshots of two colliding boson stars with $N_1 = N_2 = 10^{60}$, $\xi_1 = \xi_2 = 10^{-2}$, $\alpha = \pi$, and $v_0 = 0$, explained in section (\ref{ssec:sim3}). Contours are surfaces of constant $|\psi|^2$.}
\label{fig:Sim3}
\end{figure*}
As we can see from fig. \ref{fig:Sim3}, the two stars collide and rebound from each other. The scattering is inelastic, with translational kinetic energy being converted into vibrational energy with each collision. At much later times, the solitons will eventually merge together, as the phase of their wave functions in the overlap region rotates into some mutual value, and this has been confirmed with extended simulations.

\subsection{$N_1 = N_2$, $\xi \gg 1$, $\alpha = \pi$, $v_0 = 0$} \label{ssec:sim4}
In this case, we consider a collision between equal mass solitons out of phase with each other in the strongly-interacting regime. From our effective potential plot, we would expect the two solitons to merge easily.
\begin{figure*}
\centering
\includegraphics[width=0.9\linewidth]{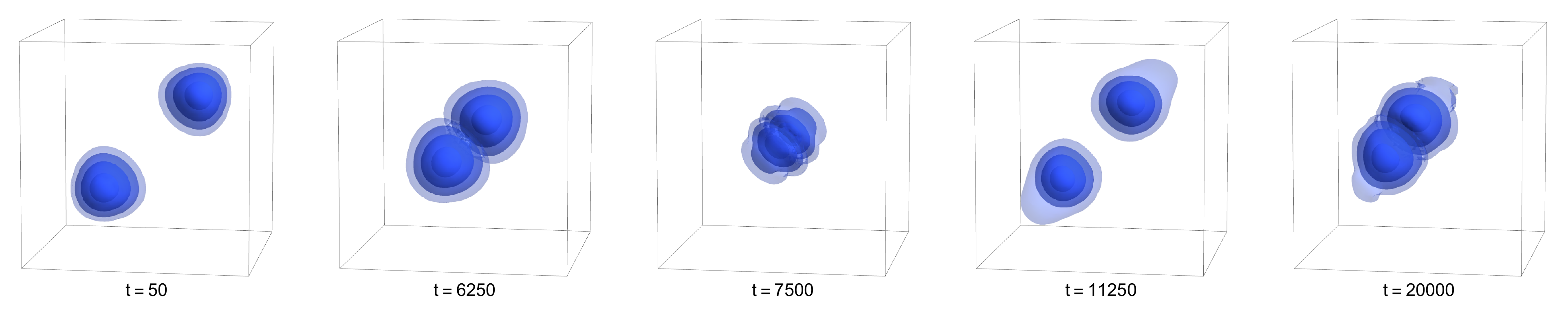}
\caption{Snapshots of two colliding boson stars with $N_1 = N_2 = 10^{60}$, $\xi_1 = \xi_2 = 10^1$, $\alpha = \pi$, $v_0 = 0$, explained in (\ref{ssec:sim4}). Contours are surfaces of constant $|\psi|^2$.}
\label{fig:Sim4}
\end{figure*}
Instead, as we can see from fig. \ref{fig:Sim4}, the solitons rebound off each other (three times in the span of this simulation before merging). The deviation from the potential prediction is likely because the potential is plotted for $\xi = 10^4$, whereas the simulation was performed with $\xi = 10^1$, so that the actual potential is more dominated by the kinetic energy contribution (which is repulsive for out of phase collisions) than the attractive effect of out of phase self-interactions, similar to the simulation of section 4.3.

\subsection{$N_1 \gg N_2$, $\xi \ll 1$, $\alpha = \pi$, $v_0 = 0$} \label{ssec:sim5}
In this case, we consider a collision between asymmetric masses in the weakly-interacting regime. The two stars are out of phase and so we expect them to repel from each other as they approach.
\begin{figure*}
\centering
\includegraphics[width=0.9\linewidth]{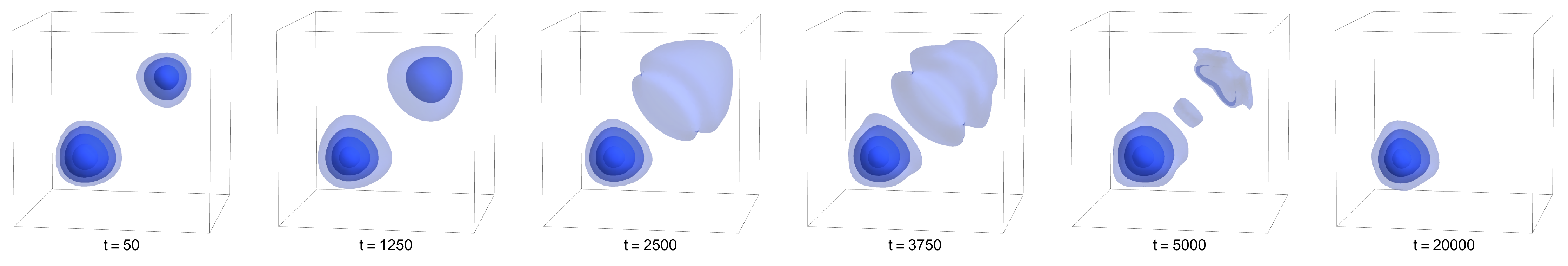}
\caption{Snapshots of two colliding boson stars with $N_1 = 10 N_2 = 10^{60}$, $\xi_1 = 10^2 \xi_2 = 10^{-2}$, $\alpha = \pi$, $v_0 = 0$, explained in (\ref{ssec:sim5}). Star one begins the simulation in the bottom left corner, and star two begins the simulation in the top right. Contours are surfaces of constant $|\psi|^2$.}
\label{fig:Sim5}
\end{figure*}
However, this prediction is dependent on the assumption that the soliton stays intact throughout the collision. As we can see from fig. \ref{fig:Sim5}, the less massive of the two stars is torn apart by tidal forces from the larger one, and material is accreted onto the star. There is mild feedback due to the fact that the two wave functions are out of phase (as seen in the rippling effect as material is siphoned from the smaller star). Since the two bodies cannot merge at once due to the phase mismatch, the smaller star must shed its mass in chunks, which then have their phase rotated to align with the phase of the larger star, and are subsequently absorbed. If there was no phase difference, this piece-by-piece accretion would not occur and would be akin to that of fig. \ref{fig:Sim2}.

\subsection{$N_1 \gg N_2$, $\xi \gg 1$, $\alpha = 0$, $v_0 = 0$} \label{ssec:sim6}
Another asymmetric-mass system, we would expect from the effective potential that the binary might be effectively stable against merger, since there is a local minimum in the energy around $x = 1.5\ell$. However, just as the situation of section 4.5, the simulation deviates from expectation, likely due to the effect of tidal forces.
\begin{figure*}
\centering
\includegraphics[width=0.9\linewidth]{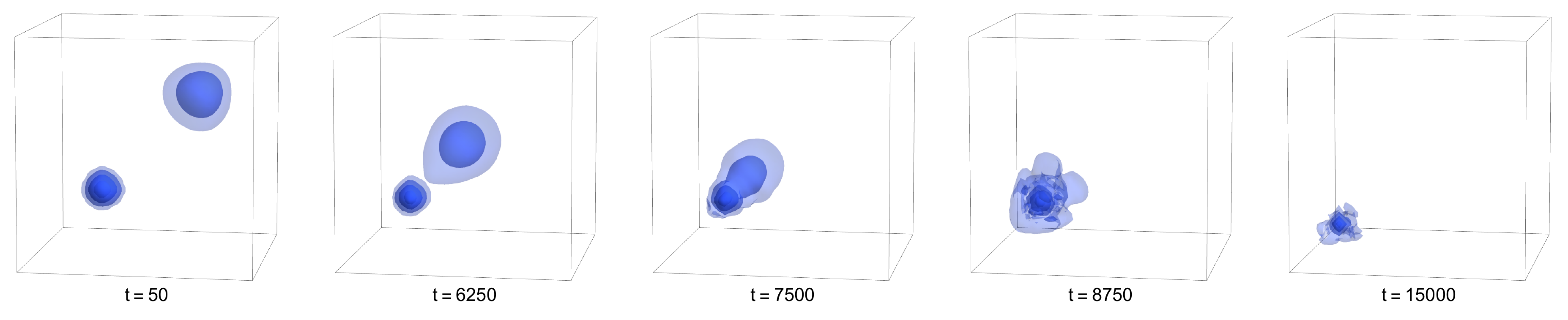}
\caption{Snapshots of two colliding boson stars with $N_1 = 4N_2 = 10^{60}$, $\xi_1 = 4^2 \xi_2 = 10^1$, $\alpha = 0$, $v_0 = 0$, explained in (\ref{ssec:sim6}). Star one begins the simulation in the bottom left corner; star two in the top right. Contours are surfaces of constant $|\psi|^2$.}
\label{fig:Sim6}
\end{figure*}
As we can see from fig. \ref{fig:Sim6}, the less massive star (upper right corner) is elongated by tidal forces and quickly accreted onto the more massive one (lower left corner). Due to the underlying grid of the simulation, especially dense configurations (such as the more massive star in this case) can get stuck on a specific grid point if the change in velocity is low enough. In this case, the more massive partner stays in the same position until the less massive partner collides with it, transferring momentum and pushing it further into the corner, as observed.

\subsection{Behavior of unstable soliton with $\xi < -1$} \label{ssec:sim7}
In this situation, we initialize a condensate in the regime $\xi < -1$, which makes the variational energy/wave vector/length scale complex, signifying instability. Since there is no variational state to initialize in, we choose the initial state wave vector to be the imaginary part of the variational wave vector (since the real part is independent of $\xi$ in this regime). We then initialize the simulation with one soliton having these parameters and allow it to evolve undisturbed.
\begin{figure*}
\centering
\includegraphics[width=0.9\linewidth]{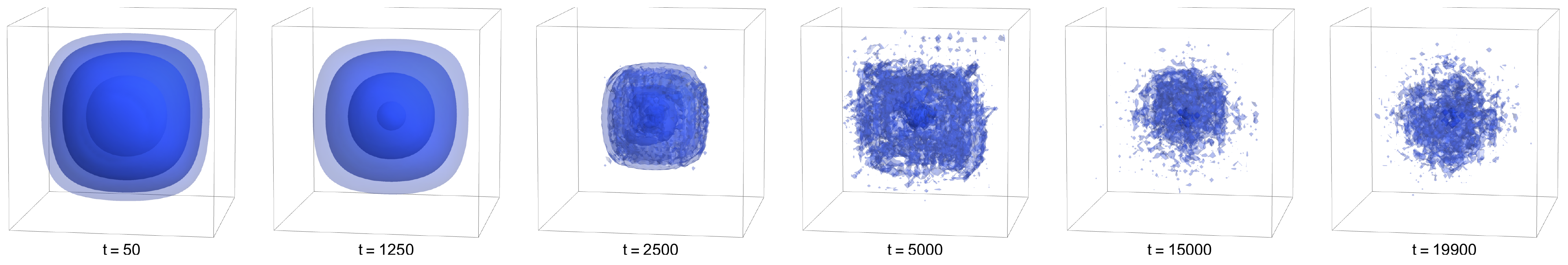}
\caption{Collapse and evolution of an unstable soliton with $N = 10^{60}$ and $\xi = -10$, explained in section (\ref{ssec:sim7}). Contours are surfaces of constant $|\psi|^2$.}
\label{fig:Sim7}
\end{figure*}
What we find is that the condensate collapses under its own gravity to an extremely dense core, with the density jumping up almost three orders of magnitude. Surrounding this core is a cloud of fluctuating points with densities roughly two orders of magnitude below the density of the core. The fact that the wave function has this discrete nature suggests large amounts of interference. This core then further fractures until it is composed of a handful of individual points of roughly the same particle number/mass.
\begin{figure*}
\centering
\includegraphics[width=0.9\linewidth]{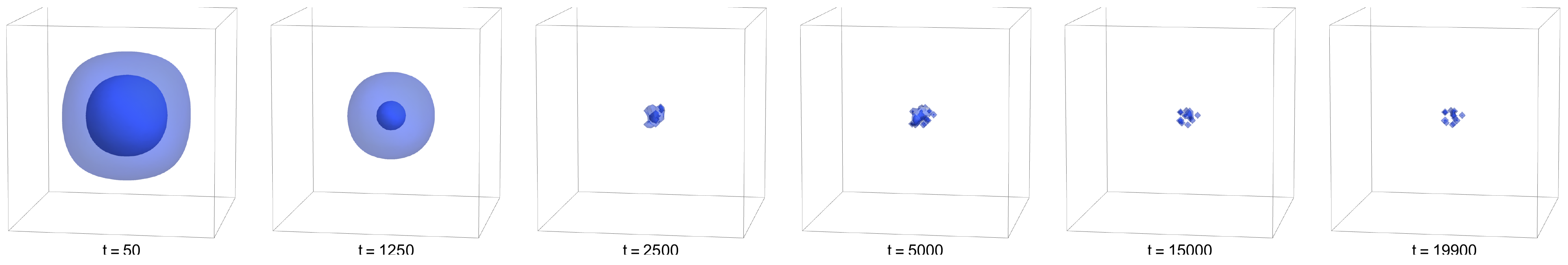}
\caption{Same simulation results as fig. \ref{fig:Sim7} from section (\ref{ssec:sim7}), but with fluctuations filtered out. Note condensate core which forms at early times $t = 2500$ further breaks up into smaller pieces as time goes on. Contours are surfaces of constant $|\psi|^2$.}
\label{fig:Sim7.2}
\end{figure*}
This fracturing of the condensate is reminiscent of the same process found in Affleck-Dine models of baryogenesis, only in this case the condensate is nonuniform and spherically symmetric to begin with, so it stands to reason it should collapse to a roughly spherical final state.

\section{Discussion} \label{Discussion}
\subsection{Results of this paper}
This study has explored the various properties of non-relativistic boson stars with a standard $\varphi^4$ self-interaction. Though this system has been studied before, both analytically and numerically, it has historically been done just for a stationary field configuration of a single soliton \cite{marpop15}\cite{penrose98}\cite{leepang92}. In recent years, some studies have looked at collisions between boson stars, though they are primarily numerical in scope \cite{bernal06}\cite{palenzuela07}\cite{palenzuela08}\cite{madarassy14}. This effective potential analysis provides good predictions for collisional behavior under a variety of circumstances and parameter ranges, and gives one the advantage of being able to understand qualitatively the outcome of a collision simply using concepts of conservation of energy. Of course, energy of translational motion of the solitons may be converted to radial oscillations, or even excite the solitons into a higher energy radial state with multiple nodes (as seen in some merger simulations), leading to dissipation of translational motion and eventual merger in most cases. Scattering from a nonzero impact parameter may lead to states with nonzero angular momenta, numerical simulations of which have been studied in \cite{palenzuela08}. Especially interesting (from an academic point of view) may be the possibility of forming stable bound states in strongly repulsive regimes, with the possibility of a centrifugal barrier provided by a nonzero angular momentum, effectively using this to construct ``molecules" from gravitationally-bound, astrophysical-scale objects. \newline
In terms of support for the predictive ability of the effective potential, it is well supported by the results of many numerical scattering simulations of boson stars. Though not all the parameter space has been adequately explored in the literature, we can make some comparisons with previous numerical studies. Simulations of self-interacting solitons with nonzero relative velocities carried out in \cite{bernal06} appear consistent with the prediction of the effective potential and the results of this studies' simulations. In Bernal and Guzm\'an's (BG) paper, they study the head-on collision of two in phase ($\alpha = 0$) boson stars with weak quartic self-interaction ($\xi \ll 1$) for both equal mass and asymmetric mass stars. The main result of the paper is that systems with ``negative" energy (systems with kinetic and self-interaction energy smaller than the gravitational binding energy) will combine and merge, whereas systems with ``positive" energy will behave solitonically and pass right through each other. This is exactly what one would predict from the effective potential, and is confirmed with simulations \ref{ssec:sim1} and \ref{ssec:sim2}. The effective potential for both mass ratios shows no ``hump" at zero separation distance (for $\alpha = 0$), and so traversing through each other is no obstacle provided there is enough translational kinetic energy to make it out of the gravitational well. As mentioned before, there are energy losses from excitation and scalar radiation during the collision, which can be treated as frictional effects, and so one cannot simply expect a system of two boson stars initially at rest with a large separation distance (such as in simulation \ref{ssec:sim1}) to have enough kinetic energy at the point of contact; an initial translational kinetic energy at infinite separation distance is required. This is also confirmed in BG's figures 4 and 5, where they explore the outcome of scattering simulations for varying initial momenta; high relative $p$ leads to solitonic behavior, whereas low $p$ leads to mergers. One downside to the effective potential formalism is that since the two stars are assumed to remain intact, it does not allow for particle transfer during the collision. This is observed in BG's paper, where scattering between asymmetric-mass boson stars appear to transfer mass between each other. \newline
The results of non-self-interacting ($\xi = 0$) boson-boson soliton scattering with different initial phases performed in \cite{palenzuela07} can be explained easily using the effective potential as well. In this paper, Palenzuela, Olabarietta, Lehner, and Liebling (POLL) use the full Einstein-Klein-Gordon system, rather than the Schr\"odinger-Poisson, though the results are quite similar. They consider scattering between both boson-boson and boson-antiboson systems with equal masses and no initial velocities for a variety of initial relative phases, though we will only compare with the boson-boson experiments, for obvious reasons. Their simulation of in-phase scattering and merger agrees with BG's and my simulations, which were discussed above. Their simulation of out-of-phase scattering, which results in the two boson stars rebounding inelastically off each other, agrees perfectly with both the effective potential prediction and the results of my simulation \ref{ssec:sim3}. The effective potential for out-of-phase systems can be seen to rise steeply as the separation distance decreases, indicating a short-range repulsive force, resulting in a gravitationally-bound state, as observed in both POLL's paper an this paper's simulation \ref{ssec:sim3}. \newline
Unfortunately, there appear to be no discussions in the literature regarding collisions between very strongly self-interacting ($\xi \gg 1$) boson stars, nor collisions between asymmetric-mass stars with $\alpha \not= 0$. Consequently, we cannot compare the results of this study to the existing literature, and it appears that this is the first attempt at simulating collisions in this parameter range. However, this is precisely this parameter range that likely requires further study, as there are multiple instances where the effective potential does not give good predictions that coincide with the outcome of the simulations. In particular, asymmetric-mass boson stars with low initial velocities are highly influenced by tidal forces, causing the smaller partner of the binary to be torn apart and accreted onto the larger partner, regardless of the value of $\alpha$. I expect that for nonzero initial velocities, where the time the two solitons spend in each other's company is reduced, the effect of tidal forces should also be reduced. The timescale of the evolution of the wave function is roughly $1/m$ and a collision that occurs on a timescale shorter than this should therefore not affect the overall shape of the boson star. As for simulations in the strongly-interacting regime, the reason the literature does not focus on this regime is likely for the same reason given in section \ref{NumericalSims}: increasing the value of $\xi$ increases the instability associated with the self-interaction term in the simulation, which requires the use of smaller and smaller time steps, making simulation in the $\xi \gg 1$ regime very computationally expensive. In addition, there is almost no discussion of boson stars with $\xi < 0$, likely due to the resultant collapse of the star.
\subsection{Implications for dark matter}
Dark matter of this composition can take on a phenomenally large array of forms, and depending on the mass of the constituent particles, can form astrophysical objects as massive as entire galaxies, and as far as I know, no lower bound on their mass exists. They can be extremely diffuse (especially in the $\xi \gg 1$ limit), leading some to interpret the galactic DM halo as a single boson star \cite{marpop15}\cite{schive14}\cite{schiveChiueh14}. This has the advantage of naturally resolving the cusp-core problem without invoking self-interactions, and their propensity to merge through low relative-velocity collisions may help explain the lack of predicted satellite galaxies. On the other hand, head-on collisions at sufficient speed can cause the DM halos to pass right through each other, as is observed in the Bullet Cluster. They can also be extremely compact, guarded against collapse to a black hole by a strong repulsive self-interaction, or if attractive, can readily collapse to a black hole at some critical mass. If this is the case, supermassive black holes at the centers of galaxies may be made of - or initially formed from - boson stars \cite{barranco11}. Somewhere between these two extremes, these configurations can form roughly stellar-mass objects that may have interesting interactions with themselves or existing stellar populations in the disk. As mentioned before, the MACHO, EROS and OGLE collaborations have all independently found a significant excess of gravitational lensing events over the number expected. Though limits have been placed on the maximum fraction of the halo mass that these objects can make up, they may still exist in a partially condensed phase, where some of the particles are contained in compact objects while others would be free-streaming. Objects on an infall trajectory from the outer parts of the halo could transfer momentum to stars and other dark objects in the galactic center through the collisional interactions elucidated in this paper, reducing the central density, as is observed in multiple DM-dominated dwarf spheroidal galaxies. Even in the non-self-interacting limit, these solitons can rebound off each other to transfer momentum and mass purely through gravity and BE statistics. Though I do not explore any of these ideas in detail in this paper, further work must be done to determine if any of these scenarios are consistent or desirable.

\section{Conclusion}
I have presented approximate analytical solutions to the nonlinear Schr\"odinger-Poisson system, constructed stable solitonic states, and explored the collisional interactions between two solitons. I have found that these boson stars can exist in a wide range of stable states, with a number of properties that make them good candidates for dark matter, such as momentum transfer (or lack thereof) through scattering and gravitational encounters, and their wide range of possible masses, densities, and length scales make them a general prediction of practically any theory of a cold, light scalar particle. I have elucidated the mechanism behind the results of direct collisions, finding it to be dependent on the relative phase, mass ratios, and self-coupling of the solitons in a way that is easily understandable in terms of energy conservation once the effective potential has been calculated. Some of these results in the weakly self-interacting regime have been confirmed by other numerical studies, while collisions with strong self-coupling and nonzero orbital angular momentum should be tested and treated in future work.

\section{Acknowledgment}
I would like to thank Alexander Kusenko for invaluable guidance while performing this work, and for useful comments during the writing/editing process; Rhyan Ghosh for helpful discussions and for keeping me focused when I didn't feel like doing any work; and the anonymous Reviewer who made many helpful suggestions regarding the presentation and scope of the paper.

\end{document}